\documentclass[twocolumn,trackchanges]{aastex7}

\hypersetup{linkcolor=red,citecolor=blue,filecolor=cyan,urlcolor=black}

\begin{document}

\title{Ultradense Dark Matter Halos with Poisson Noise from Stellar-Mass Primordial Black Holes} 

\author[orcid=0000-0002-6349-8489]{Saeed Fakhry} 
\affiliation{Department of Physics, K.N. Toosi University of Technology, P.O. Box 15875-4416, Tehran, Iran}
\email[show]{s\_fakhry@kntu.ac.ir} 

\author[orcid=0000-0003-3229-3429]{Javad T. Firouzjaee} 
\affiliation{Department of Physics, K.N. Toosi University of Technology, P.O. Box 15875-4416, Tehran, Iran}
\email[show]{firouzjaee@kntu.ac.ir}

\begin{abstract}
\noindent
In this work, we investigate the impact of Poisson noise from stellar-mass primordial black holes (PBHs) on the formation of ultradense dark matter halos (UDMHs). Our findings reveal that the discrete spatial distribution of PBHs significantly enhances small-scale density fluctuations, particularly for massive stellar-mass PBHs. Our results indicate that the modified power spectrum, incorporating both adiabatic and isocurvature contributions from PBH-induced Poisson noise, strongly depends on PBH mass and fraction. Specifically, increasing PBH mass shifts the differential mass function of UDMHs toward higher masses, while variations in the suppression parameter $n$ modulate the efficiency of UDMH formation at small scales. For lower values of $n$, our findings show a significant boost in UDMH abundance, favoring multi-component dark matter scenarios. Conversely, at higher values of $n$, the predicted UDMH distributions align more closely with single-component models dominated by stellar-mass PBHs. Furthermore, our analysis demonstrates that more realistic halo mass functions, which account for angular momentum and dynamical friction, consistently predict higher UDMH abundances compared to traditional Press-Schechter formalism.
\end{abstract}


\section{Introduction} 
The hierarchical nature of cosmic structure formation is characterized by a bottom-up evolutionary path, in which quantum fluctuations in the scalar field were stretched to cosmological scales, generating primordial density perturbations that could lead to the formation of gravitationally bound objects \citep{1972MNRAS.160P...1Z, 1982PhLB..117..175S, 1982PhRvL..49.1110G, 1984PThPS..78....1K, 1992PhR...215..203M}. While this process is largely driven by the initial perturbations of moderate amplitude from inflation ($\delta \sim 10^{-5}$), there may be some theoretical frameworks that predict deviations from this conventional paradigm; \citep[see, e.g.,][]{1992JETPL..55..489S, 2000PhRvD..62d3508C, 2001PhRvD..63l3501M, 2008PhRvD..77b3514J, 2010AdAst2010E..76B, 2010CQGra..27l4010K, 2011JCAP...05..024A}. In this regard, primordial black holes (PBHs) are strong candidates for macroscopic dark matter, potentially formed through rare, large-amplitude ($\delta \gtrsim 0.3$) density fluctuations in the early Universe; \citep[see, e.g.,][]{2021FrASS...8...87V, 2021JPhG...48d3001G, 2021arXiv211002821C, 2023JCAP...05..054K}. Over the past few years, a variety of dark matter halo models have been proposed to describe the gravitational waves observed by the LIGO-Virgo-KAGRA detectors within the framework of the PBH scenario and to impose constraints on the abundance of PBHs as a constituent of dark matter; \citep[see, e.g.,][]{2021PhRvD.103l3014F, 2022PhRvD.105d3525F, 2022ApJ...941...36F, 2023PDU....4101244F, 2023PhRvD.107f3507F, 2023ApJ...947...46F, 2023arXiv230811049F, 2024arXiv240115171F, 2024ApJ...976..248F}. Moreover, it has recently been shown that primordial non-singular black holes can account for all dark matter within a significantly broader parameter space compared to Schwarzschild PBHs \citep{2025PhRvD.111b4009C, 2025PhRvD.111b4010C}.

An essential but often overlooked component of PBH-dominated dark matter scenarios is the intrinsic Poisson statistics that dictate their discontinuous distribution \citep{2003ApJ...594L..71A}. The formation of PBHs induces entropy perturbations that appear in the matter power spectrum and give rise to a Poisson spatial distribution \citep{2019JCAP...11..015A}. This feature significantly distinguishes PBH models from particle dark matter frameworks, where the matter distribution is usually approximated as a continuous fluid \citep{2018JCAP...07..005O}. The inherent discreteness of PBH distributions introduces significant shot noise, which substantially alters the matter power spectrum at small scales and, consequently, the dynamics of the formation of compact dark matter structures \citep{2021arXiv211110028W}. These Poisson fluctuations affect not only the direct distribution of PBHs, but also the formation of other dark matter structures, especially in moderate-amplitude overdensities. The magnitude of this shot noise contribution scales inversely with the number density of PBHs, making it particularly relevant for scenarios involving massive PBHs where their spatial number density is naturally lower \citep{2017PDU....18...47G}. 

Furthermore, in contrast to conventional cold dark matter predictions, Poisson fluctuations may accelerate the growth of cosmic structures in the early Universe by inducing early structure formation. In regions where the characteristic scale of the shot noise approaches that of the primordial disturbances, the interaction between Poisson noise and the underlying density field becomes particularly complex, resulting in non-trivial modifications of the normal structure development paradigm \citep{2015APh....63...23H}. Given that next-generation surveys investigating the high-redshift Universe may soon verify these predictions, it is imperative to comprehend these alterations in order to make accurate predictions about the quantity and characteristics of early cosmic structures in PBH scenarios \citep{2024SCPMA..6709512Y, 2024PhRvD.110j3540H}.

Although PBHs emerge from substantial density fluctuations, more moderate overdensities ($\delta\gtrsim 10^{-3}$) can facilitate the formation of ultradense dark matter halos (UDMHs) \citep{2009PhRvL.103u1301S, 2009ApJ...707..979R, 2023MNRAS.520.4370D, 2024PDU....4601544F}. These structures, while lacking an event horizon, exhibit greater sensitivity to small-scale perturbations, thereby acting as sensitive probes of the initial small-scale power spectrum \citep{2012PhRvD..85l5027B}. The formation of dark matter halos has been investigated using both numerical simulations and analytical methods, with a particular focus on those that emerged during the matter-dominated era, \citep[see, e.g.,][]{2017PhRvD..96l3519G, 2018PhRvD..98f3527D, 2022MNRAS.517L..46W, 2023PDU....4101259D, 2024ARep...68...19D}. Numerous studies have been conducted to specifically explore the formation of UDMHs during the radiation-dominated era, \citep[see,][]{1994PhRvD..50..769K, 2002JETP...94....1D, 2010PhRvD..81j3529B, 2013JCAP...11..059B, 2019PhRvD..99l3530N}. The mass scale of UDMHs is intrinsically linked to the horizon mass at the time fluctuations re-enter the cosmological horizon. Smaller-scale perturbations, which re-enter earlier, experience extended periods of overdensity growth, enhancing their development. The interplay between Poisson-induced fluctuations and these moderate-density regions can establish distinct conditions for UDMH formation, setting them apart from conventional structure formation processes.

In this study, we introduce a modified matter power spectrum that incorporates the influence of Poisson noise arising from stellar-mass PBHs and examine its implications for the formation of UDMHs. The paper is structured as follows: In Section \ref{sec:ii}, we analyze the formation and evolution of UDMHs during the radiation-dominated epoch. Additionally, we characterize their abundance using more realistic mass functions and a refined power spectrum that accounts for Poisson noise contributions from stellar-mass PBHs. In Section \ref{sec:iii}, we present our results and provide a detailed discussion on the abundance of UDMHs. Finally, in Section \ref{sec:iv}, we summarize the key conclusions of our investigation.
\section{Formation of UDMHs}\label{sec:ii}
The evolution of linear density perturbations in dark matter during the radiation-dominated era, initiated by primordial curvature perturbations $\zeta$, is characterized by:
\begin{eqnarray}\label{eq1}
\delta(k, a) = \alpha \zeta(k) \log\left(\beta\frac{a}{a_{\rm H}}\right) \hspace{1.2cm} \nonumber \\
=\alpha \zeta(k)\log\left(\sqrt{2}\beta\frac{k}{k_{\rm eq}}\frac{a}{a_{\rm eq}}\right),
\end{eqnarray}
where $a \gg a_{\rm H}$, with $a_{\rm H}$ being the scale factor at horizon entry. The parameters $k_{\rm eq} \simeq 0.01\,{\rm Mpc^{-1}}$ and $a_{\rm eq} \simeq 3\times 10^{-4}$ represent the horizon scale and scale factor at matter-radiation equality, respectively. Numerical estimates suggest $\alpha = 6.4$ and $\beta = 0.47$ \citep{1996ApJ...471..542H}. Additionally, the scale factor at horizon crossing is given by:
\begin{eqnarray}
\frac{a_{\rm H}}{a_{\rm eq}}=\frac{1+\sqrt{1+8(k/k_{\rm eq})^{2}}}{4(k/k_{\rm eq})^{2}}
\simeq \frac{\sqrt{2}}{2}\frac{k_{\rm eq}}{k}, \hspace*{0.4cm} k\gg k_{\rm eq}.
\end{eqnarray}

Non-relativistic dark matter is thought to have been disconnected from radiation during the horizon crossing. Individual dark matter particles' motion, driven by curvature perturbations, can be described as an ellipsoid's axis drifting independently. The initial tidal field of an area with scale $k$ dictates its eccentricity $e$ and prolateness $p$, which in turn influence the axis ratios of the ellipsoidal drift. The density evolution in such an area depends on the contraction or expansion of each axis caused by particle motion, as defined by:
\begin{equation}
\frac{\rho}{\bar{\rho}_{\rm m}} = \prod_{i=1}^{3} \left|1 - \lambda_{i}\delta(k, a)\right|^{-1},
\end{equation}
where $\bar{\rho}_{\rm m}$ denotes the mean dark matter density, and $\lambda_{i}$ are the eigenvalues of the deformation tensor under the Zeldovich approximation:
\begin{equation}
\lambda_{1}=\frac{1+3e+p}{3}, \hspace*{0.2cm}\lambda_{2}=\frac{1-2p}{3}, \hspace*{0.2cm}\lambda_{3}=\frac{1-3e+p}{3}.
\end{equation}
Collapse along a given axis $i$ occurs when the linear density perturbation $\delta(k, a)$ exceeds $\lambda_{i}^{-1}$. The axis associated with the smallest eigenvalue ($\lambda_{3}$) collapses last, and the entire ellipsoidal region collapses when $\lambda_{3}\delta(k, a_{\rm c}) = 1$, where $a_{\rm c}$ denotes the collapse scale factor. The critical density contrast under these conditions is given by:
\begin{equation}\label{eq_delta}
\delta_{\rm c} = \frac{3}{1-3e+p}.
\end{equation}
The most likely values for $e$ and $p$ are $e_{\rm mp} = \sigma/(\sqrt{5}\delta)$ and $p_{\rm mp} = 0$, respectively, given a Gaussian random field with density contrast $\delta$ and linear root-mean-square $\sigma$ \citep{2001MNRAS.323....1S}. According to the excursion set formalism, the moving barrier is $B(S) = 3(1+\sqrt{S/5})$, with $S \equiv \sigma^{2}$. These values result in a collapse threshold of $\delta_{\rm c} = 3(1+\sigma/\sqrt{5})$ \citep{1991ApJ...379..440B}. Therefore, collapse can happen during the radiation-dominated epoch if the density perturbations are high enough. Only the rarest spikes can produce PBHs, although moderate perturbations can make UDMHs.

It is crucial to note that halos did not always form from collapsing structures during the radiation-dominated era. Numerical simulations show that a collapsed peak only creates a virialized halo when it changes to a locally matter-dominated state \citep{2019PhRvD.100j3010B}. In contrast, collapsed regions are substantially denser than the surrounding dark matter, with their densities scaling as $\sim e^{-2}\bar{\rho}_{\rm m}$ \citep{2023MNRAS.520.4370D}. Given the typical ellipticity $e\sim0.15$ of the tidal field at a $3\sigma$ peak, local matter dominance occurs at $a/a_{\rm eq} \sim e^{2} \simeq \mathcal{O}(10^{-2})$. As a result, halos form long before matter dominion begins. The density within these halos is quite high, as they scale with the average cosmic density at the time of formation

\subsection{Halo mass functions}
The unconditional halo mass function, which characterizes the average comoving number density of halos inside a logarithmic mass interval, is provided by the excursion set theory \citep{1991ApJ...379..440B}.
\begin{equation}
\frac{{\rm d}n}{{\rm d}\log M} = \frac{\bar{\rho}_{\rm m,0}}{M} \left|\frac{{\rm d}\log\nu}{{\rm d}\log M} \right| \nu f(\nu).
\end{equation}
In this case, the comoving average density of dark matter content is represented as $\bar{\rho}_{\rm m,0} \simeq 33\, {\rm M_{\odot}/kpc^{3}}$, and the dimensionless parameter $\nu \equiv \delta_{\rm c}/\sigma$ is referred to as peak height. Additionally, the distribution of first crossings is represented by the function $f(\nu)$, also referred to as the "multiplicity function." The root-mean-square of linear density fluctuations smoothed on the mass scale $M$ is represented by:
\begin{equation}
\sigma^2(M, a) = \frac{1}{2\pi^{2}}\int_{0}^{\infty} P(k, a) W^2(k, M) k^{2} {\rm d}k.
\end{equation}
The smoothing window function $ W(k, M) $, which is considered to be a sharp-$k$ filter \footnote{The sharp-$k$ filter is chosen for its effectiveness in analyzing the small-scale power spectrum, the focus of this study. Unlike a real-space top-hat window function, which introduces variance and noise due to its oscillatory Fourier transform, the sharp-$k$ filter ensures a clean truncation in Fourier space, minimizing artificial oscillations. While Gaussian or top-hat functions are common, they either blur scale distinctions (Gaussian) or increase variance (top-hat), making them less suitable for isolating small-scale effects \citep{2020JCAP...12..038T, 2022ApJ...928L..20S, 2023A&A...669L...2A}.}, is defined as $ W(k, M) = 1 $ if $ 0 < k \leq k_M $, and $ W(k, M) = 0 $ otherwise, where $ k_M = \left(6\pi^2 \bar{\rho}_{\rm m,0} / M\right)^{1/3} $ \citep{1994MNRAS.271..676L}. Additionally, $ P(k, a) $ represents the power spectrum of linear matter fluctuations \citep{1991ApJ...379..440B}. 
In these situations, one can have
\begin{equation}
\frac{{\rm d}\log \nu}{{\rm d}\log M}=\frac{k_{M}^{3}}{12\pi^{2}}\frac{P(k_{M}, a)}{\sigma^{2}(M, a)}.
\end{equation}
From Eq.\,(\ref{eq1}), it follows that
\begin{equation}
P(k, a)=\frac{2\pi^{2}\alpha^{2}}{k^{3}}\left[\log\left(\sqrt{2}\beta \frac{k}{k_{\rm eq}}\frac{a}{a_{\rm eq}}\right)\right]^{2}P_{\zeta}(k),
\end{equation}
where $P_{\zeta}(k)$ represents the dimensionless power spectrum of primordial curvature perturbations. The exact shape of $P_{\zeta}(k)$ depends on the hypothesized inflationary scenario that causes the perturbations. In the next section, we propose modifying the power spectrum to account for the Poisson noise of stellar-mass PBHs at small scales.

This study aims to investigate how physical elements affect the halo mass function in the ultradense domain of excursion set theory, specifically their impact on the dynamical barrier. To determine the distribution of initial crossings in scenarios with various barriers, simulate a large number of random walks \citep{1998MNRAS.300.1057S, 2002MNRAS.329...61S}. According to \cite{2002MNRAS.329...61S}, for various dynamical barriers, the initial crossing distribution can be approximated by
\begin{equation} 
f(S) = |T(S)|\exp\left(-\frac{B(S)^{2}}{2S}\right)\frac{1}{S\sqrt{2\pi S}},
\end{equation}
where $T(S)$ is given by the Taylor expansion of $B(S)$
\begin{equation}
T(S) = \sum_{n=0}^{5}\frac{(-S)^{n}}{n!}\frac{\partial^{n}B}{\partial S^{n}}.
\end{equation}
\cite{2023MNRAS.520.4370D} derive an approximation for the distribution of first barrier crossings within a Gaussian random walk framework, validated by Monte Carlo simulation,
\begin{equation}
f(S) \approx \frac{3 + 0.556\sqrt{S}}{\sqrt{2\pi S^{3}}} \exp\left(-\frac{B^2}{2S}\right) \left(1 + \frac{S}{400}\right)^{-0.15},
\end{equation}
which corresponds to the PS multiplicity function\footnote{The statement explains how the multiplicity function $f(\nu)$ is connected to the first crossing distribution $f(S,t)$. The latter represents the probability distribution for the first time a random walk crosses a specific barrier $B(S)$. The equation $\nu f(\nu) = Sf(S,t)$ demonstrates that $f(\nu)$ can be derived from $f(S,t)$, illustrating a fundamental relation within the excursion set theory \citep{2007IJMPD..16..763Z}.},
\begin{equation}
\left[\nu f(\nu)\right]_{\rm PS} = \sqrt{\frac{2}{\pi}}\frac{(\nu + 0.556) \exp\left[-0.5(1+\nu^{1.34})^2\right]}{(1+0.0225\nu^{-2})^{0.15}}.
\end{equation}

It should be noted that discrepancies exist between the predictions of the PS mass function and the distribution of dark matter halos, particularly at high redshifts \citep{2017JCAP...03..032D}. These discrepancies, as previously noted, may arise from various physical factors overlooked in the PS formalism, which could significantly influence the abundance of UDMHs. One key factor addressed here involves geometry, extending the spherical-collapse model foundational to the PS formalism to ellipsoidal-collapse halo models. This modification leads to the Sheth-Tormen (ST) mass function, as outlined by \cite{2001MNRAS.323....1S}
\begin{equation}
\left[\nu f(\nu) \right]_{\rm ST} = A_{1}\sqrt{\frac{2 \nu^{\prime}}{\pi}}\left(1+\frac{1}{\nu^{\prime q}}\right)\exp\left(-\frac{\nu^{\prime}}{2}\right),
\end{equation}
where $q=0.3$, $\nu^{\prime} \equiv 0.707 \nu^{2}$, and $A_{1}=0.322$ determined by ensuring that the integral of $f(\nu)$ over all possible values of $\nu$ equals unity.

In addition to the geometric parameters affecting the virialization of dark matter halos, certain physical factors also substantially influence the collapse of overdense regions and, as a result, the halo mass function. Including these critical physical factors in the analysis is essential, as they encapsulate the fundamental physics driving halo collapse and growth, as well as the mechanisms underlying the formation and evolution of cosmic structures over time. This methodology enables the collapse threshold to dynamically depend on effective physical parameters. Consequently, the barrier adapts to these variables, yielding a more accurate model for halo collapse. Key considerations in these corrections include the effects of angular momentum, dynamical friction, and the cosmological constant on the halo mass function. These adjustments help mitigate discrepancies, particularly in contested mass ranges. Incorporating angular momentum and the cosmological constant yields a revised mass function, known as DP1 in this analysis \citep{2006ApJ...637...12D}
\begin{equation}
\left[\nu f(\nu)\right]_{\rm DP1} \approx A_{2} \sqrt{\frac{\nu^{\prime}}{2\pi}} k(\nu^{\prime}) \exp\{-0.4019 \nu^{\prime}l(\nu^{\prime})\},
\end{equation}
with $A_{2} = 0.974$ established through normalization, and
\begin{equation}
k(\nu^{\prime}) = \left(1 + \frac{0.1218}{(\nu^{\prime})^{0.585}} + \frac{0.0079}{(\nu^{\prime})^{0.4}} \right),
\end{equation}
and
\begin{equation}
l(\nu^{\prime}) = \left(1 + \frac{0.5526}{(\nu^{\prime})^{0.585}} + \frac{0.02}{(\nu^{\prime})^{0.4}} \right)^{2}.
\end{equation}

The impact of dynamical friction on the barrier was also examined in \cite{2017JCAP...03..032D}, resulting in the mass function referred to as DP2 hereafter
\begin{equation}
\left[\nu f(\nu)\right]_{\rm DP2} \approx A_{3} \sqrt{\frac{\nu^{\prime}}{2\pi}}m(\nu^{\prime}) \exp\{-0.305 \nu^{\prime 2.12}n(\nu^{\prime})\},
\end{equation}
where $A_{3} = 0.937$ is determined through normalization, and we obtain
\begin{equation}
m(\nu^{\prime}) = \left(1 + \frac{0.1218}{(\nu^{\prime})^{0.585}} + \frac{0.0079}{(\nu^{\prime})^{0.4}} + \frac{0.1}{(\nu^{\prime})^{0.45}}\right),
\end{equation}
and
\begin{equation}
n(\nu^{\prime}) = \left(1 + \frac{0.5526}{(\nu^{\prime})^{0.585}} + \frac{0.02}{(\nu^{\prime})^{0.4}} + \frac{0.07}{(\nu^{\prime})^{0.45}}\right)^{2}.
\end{equation}
In Section~\ref{sec:iii}, we will employ these mass functions to compute the abundance of UDMHs for the modified power spectrum while incorporating Poisson noise contributions from stellar-mass PBHs.
\subsection{Modified power spectrum}
The formation of PBHs and UDMHs in the radiation-dominated era demands a significant amplification of the primordial power spectrum at small scales, which contrasts with the nearly scale-invariant spectrum observed at larger scales. Hence, it becomes imperative to refine the scale-invariant power spectrum through the incorporation of large-amplitude small-scale fluctuations, as such modifications can enhance the probability of PBH formation.

\begin{figure*}
\centering
\includegraphics[width=1\textwidth]{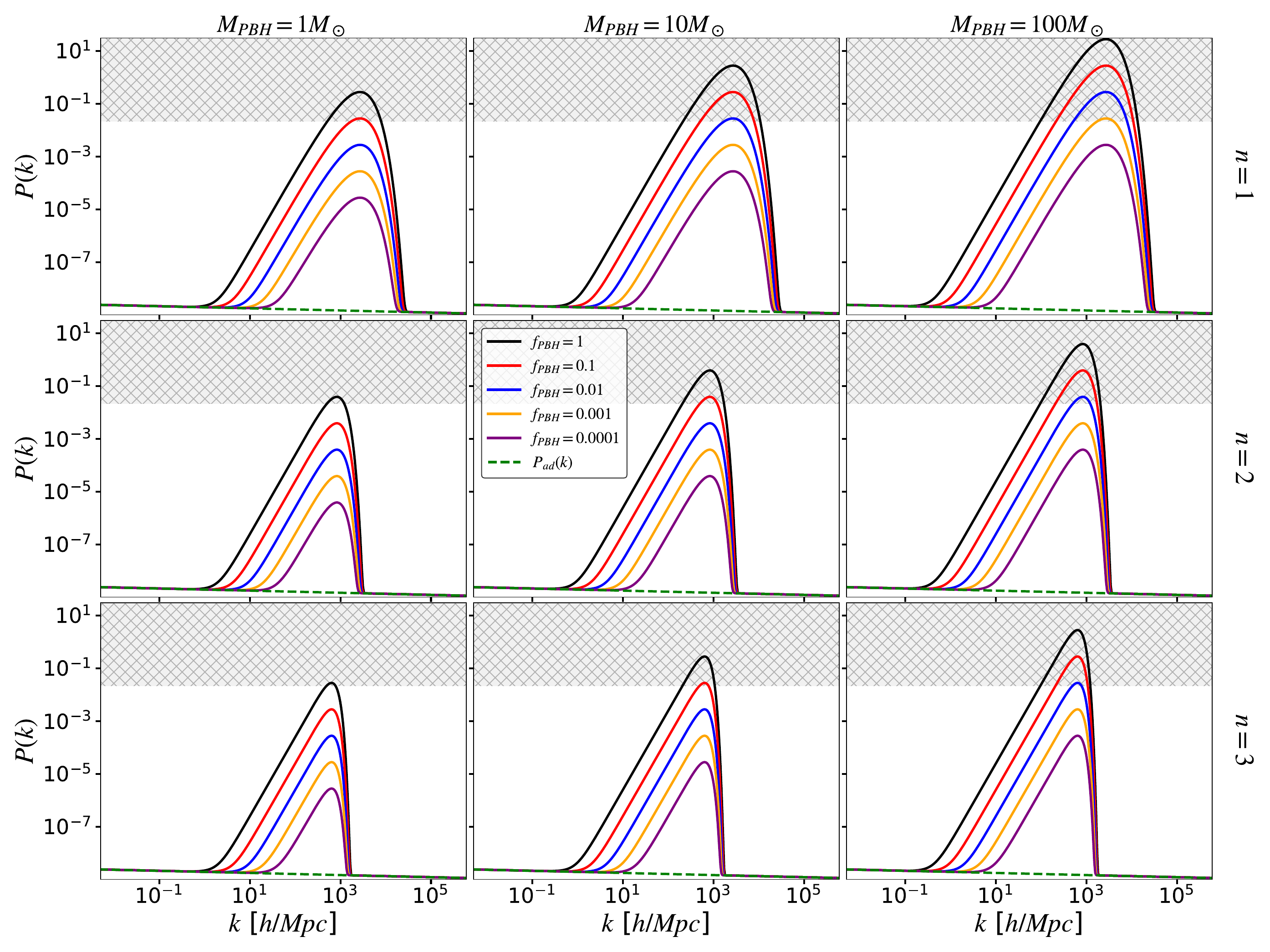}
\caption{Modified matter power spectrum as a function of wavenumber $k$, described by Eq.\,(\ref{powerspec1}), while incorporating the impact of Poisson noise from stellar-mass PBHs in the mass range of $M_{\rm PBH}=1, 10$ and $100 M_{\odot}$, for various values of the fraction $f_{\rm PBH}$. Also, the suppression parameter is considered to be $n=1, 2,$ and $3$.} The shaded gray region indicates the range where the amplitude of the power spectrum on small scales is enhanced by a factor of at least seven orders of magnitude relative to its value on large scales.
\label{Fig1}
\end{figure*}

We analyze the formation of PBHs as a consequence of a phase transition occurring at a critical temperature $T_{\rm c}$ within the radiation-dominated era \citep{2002PhRvD..66f3505B}. The emergence of PBHs is hypothesized to result from quantum fluctuations or phase transitions in the early Universe, which induce the collapse of overdense regions into black holes. In this context, deviations from scale invariance, particularly through enhanced small-scale fluctuations, can significantly impact the abundance and mass distribution of PBHs. Under the assumption that accretion following formation is negligible, the black hole mass is considered to be approximately $M_{\text{PBH}}$. In this regard, the energy density associated with PBHs can be specified as
\begin{equation}\label{rhopnh}
\rho_{\text{PBH}} = M_{\text{PBH}}\,n_{\text{PBH}} (1 + \delta_{\rm p}) \left( \frac{T}{T_{\rm c}} \right)^3,
\end{equation}
where $n_{\text{PBH}}$ denotes the comoving number density of PBHs, $\delta_{\rm p}$ represents the Poisson fluctuations inherent in their distribution, and the factor $(T/T_c)^3$ accounts for the dilution in PBH density due to cosmic expansion. Given that $n_{\text{PBH}}$ is a conserved quantity, one can employ the relation $M_{\rm PBH} n_{\rm PBH} = f_{\rm PBH} \Omega_{\rm CDM}\rho_{\rm c}$, where $f_{\rm PBH}=\Omega_{\rm PBH}/\Omega_{\rm CDM}$ is the fraction of PBHs in the cold dark matter content, and $\rho_{\rm c} = 2.78 \times 10^{11} h^2 \, M_{\odot} \, \mathrm{Mpc}^{-3}$ represents the present-day critical density.

Primordial black holes do not exhibit correlations on scales larger than the Hubble horizon. In this context, the corresponding power spectrum is given by \citep{2003ApJ...594L..71A}:
\begin{equation}
P_{\rm p} = \langle |\delta_{\rm p}(k)|^2 \rangle \simeq n_{\text{PBH}}^{-1}.
\end{equation}
Under such assumptions, the power spectrum of primordial curvature perturbations can be determined as:
\begin{eqnarray} \label{powerspec1}
P(k) = P_{\rm ad}(k) + P_{\rm iso} (k),
\end{eqnarray}
where $P_{\rm ad}(k)$ represents the adiabatic power spectrum generated by inflation (or another approach to producing scale-invariant adiabatic perturbations):
\begin{eqnarray}
P_{\rm ad}(k)= A_{\rm s}\left(\frac{k}{k_{\rm s}}\right)^{n_{\rm s}-1}.
\end{eqnarray}
Here $k_{\rm s}=5\times 10^{-2}\,\mathrm{Mpc}^{-1}$ is the pivot scale and $A_{\rm s} = 2.1 \times 10^{-9}$ and $n_{\rm s} = 0.96$ are set from cosmic microwave background observations \citep{2020A&A...641A...6P}. Moreover, $P_{\rm iso}(k)$ is the contribution of Poisson noise from PBHs, which can be defined as
\begin{equation}\label{eqn}
P_{\rm iso}(k) = Q(k)\,\exp\left\{\frac{-[(k-k^{\prime})/\sigma]^{n}}{2}\right\}.
\end{equation}
In the above relation 
\begin{eqnarray}
Q(k) = A_{\rm iso}\left(\frac{k}{k_{\rm s}}\right)^{n_{\rm iso}-1},
\end{eqnarray}
where
\begin{eqnarray} 
A_{\rm iso} = 3.2\times 10^{-12} f_{\rm PBH} \left(\frac{M_{\rm PBH}}{30 M_{\odot}}\right),
\end{eqnarray} 
is the amplitude of the isocurvature perturbation at the pivot scale, and $n_{\rm iso}=4$ \citep{2017JCAP...08..017G}. Furthermore, an exponential function is employed to suppress amplified fluctuations on small scales beyond the cut-off wavenumber $k^{\prime} \simeq 10^{2}\,{\rm Mpc^{-1}}$, with a characteristic width of $\sigma \simeq 4.5 \times 10^{2}\,{\rm Mpc^{-1}}$. Also, $n$ is the parameter that governs the amount of suppression. Also, $n$ is the parameter that governs the amount of suppression, which plays a crucial role in regulating the steepness of the exponential cutoff in the isocurvature power spectrum induced by Poisson fluctuations from PBHs. Physically, this parameter encodes the nature of small-scale damping mechanisms that can emerge from nonlinear effects, thermal diffusion, or mode-mode coupling in the early Universe. While the exact microphysical origin of this suppression remains model-dependent, a phenomenological exponential cutoff has been widely adopted in the literature to model abrupt decreases in power beyond a characteristic wavenumber, \citep{2017JCAP...08..017G}. Moreover, we focus on the Poisson noise from stellar-mass PBHs. Consequently, the mass range of PBHs is chosen to span $M_{\rm PBH}=(1\mbox{-}100) M_{\odot}$.

In Fig.\,\ref{Fig1}, we have plotted the modified matter power spectrum as a function of wavenumber $k$, incorporating the impact of Poisson noise from stellar-mass PBHs across different mass ranges $M_{\rm PBH}=1, 10$, and $100 M_{\odot}$, while varying the suppression parameter $n=1, 2, 3$ and the PBH fraction $f_{\rm PBH}=1, 0.1, 0.01, 0.0001$, and $0.0001$. The selected range $n = 1$ to $3$ spans the transition from a relatively mild cutoff (broad enhancement across small scales) to a sharper suppression (localized enhancement near $k \sim 100\,\text{Mpc}^{-1}$. This approach allows us to explore the sensitivity of UDMH formation to varying suppression strengths while remaining agnostic about the exact microphysical damping source. The figure demonstrates how the power spectrum is significantly enhanced at small scales compared to large scales, with the shaded region indicating an amplification factor of at least seven orders of magnitude. This enhancement is crucial for understanding the formation of UDMHs and PBHs during the radiation-dominated era. The different panels represent varying degrees of suppression of amplified fluctuations beyond the cutoff wavenumber, showing how increasing the suppression parameter $n$ leads to progressively steeper declines in power at higher wavenumbers, thus moderating the influence of Poisson noise on smaller scales.

The figure reveals that the impact of Poisson noise becomes increasingly pronounced as PBH mass increases from $1$ to $100 M_{\odot}$, highlighting the complex interplay between noise-induced fluctuations and underlying density fields. For lighter PBHs ($1 M_{\odot}$), the modifications to the power spectrum remain relatively modest even as $f_{\rm PBH}$ varies, while heavier PBHs ($100 M_{\odot}$) introduce substantial changes, particularly at small scales. This behavior can be attributed to the inverse relationship between PBH number density and mass, where lower number densities of massive PBHs result in stronger shot noise effects. Additionally, varying $f_{\rm PBH}$ significantly affects the overall amplitude of the power spectrum; higher $f_{\rm PBH}$ values lead to more substantial modifications, underscoring the sensitivity of structure formation processes to the relative abundance of PBHs within the dark matter content.

\begin{figure*}
\includegraphics[width=\textwidth]{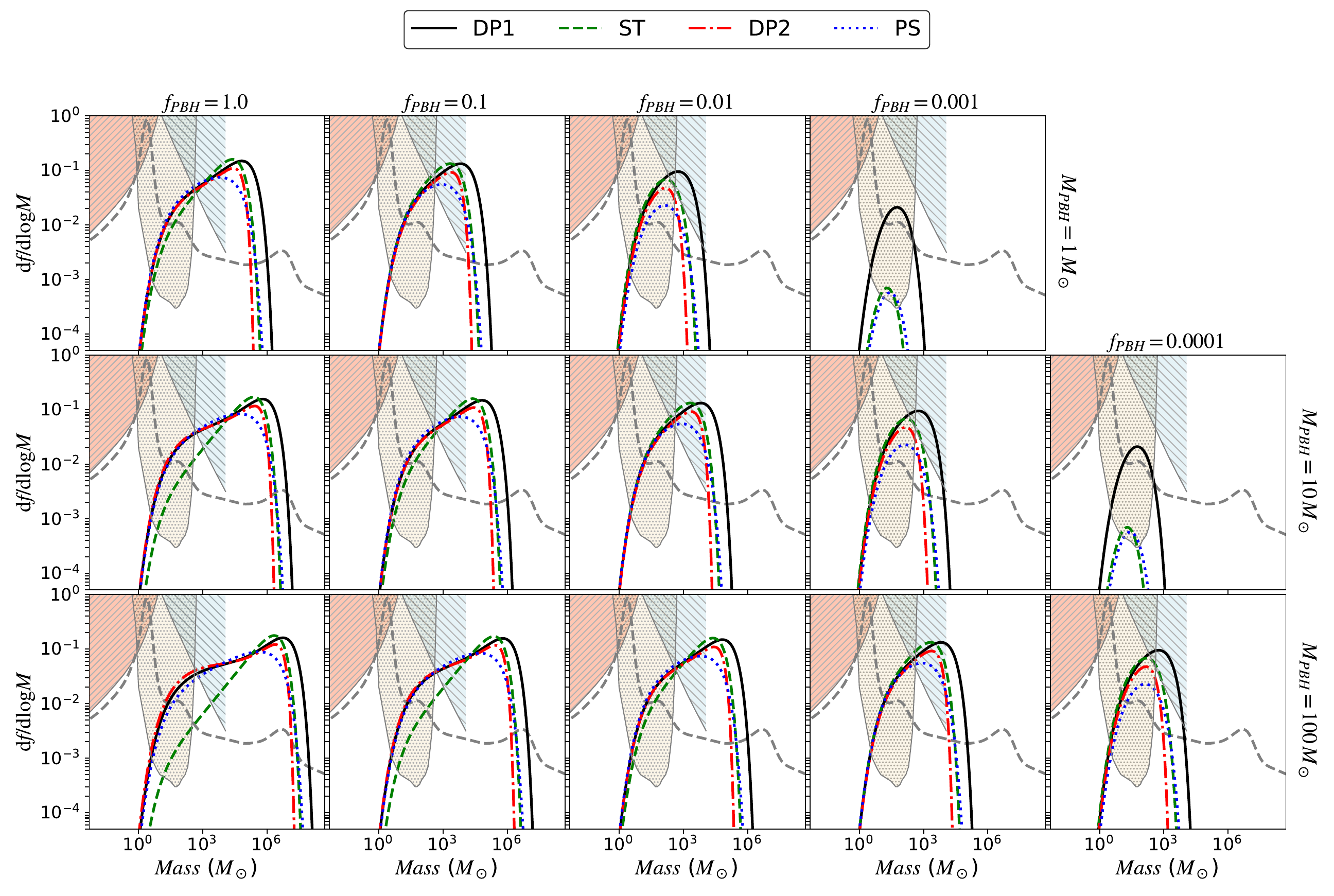}
\caption{The differential mass fraction of dark matter in UDMHs is expressed as a function of mass $M$, incorporating the modified power spectrum derived in Eq.~(\ref{powerspec1}). This analysis accounts for Poisson noise contributions from PBHs with masses $M_{\rm PBH}=1,10$, and $100 M_{\odot}$, while considering multiple values of the PBH fraction $f_{\rm PBH}$, and assumes a spectral index $n=1$. The results for different mass functions are illustrated using various line styles: black solid for DP1, green dashed for ST, red dot-dashed for DP2, and blue dotted for PS at $a = a_{\rm eq}$. Additionally, the gray dashed line represents the PBH mass spectrum derived from the thermal history of the early Universe \citep{2019arXiv190608217C}. The upper limits from observational constraints on PBHs are included. These constraints are based on microlensing data from OGLE \citep{2024ApJ...976L..19M}, direct constraints on PBH mergers from LVK observations \citep{2019PhRvD.100b2017A, 2022PhRvL.129f1104A, 2024PhRvD.110b3040A}, and the disruption of ultra-faint dwarfs (UFD) \citep{2016ApJ...824L..31B}, that are presented as shaded orange, wheat and cyan areas with their own unique hatching, respectively.} \label{Fig2}
\end{figure*}

\begin{figure*}
\includegraphics[width=\textwidth]{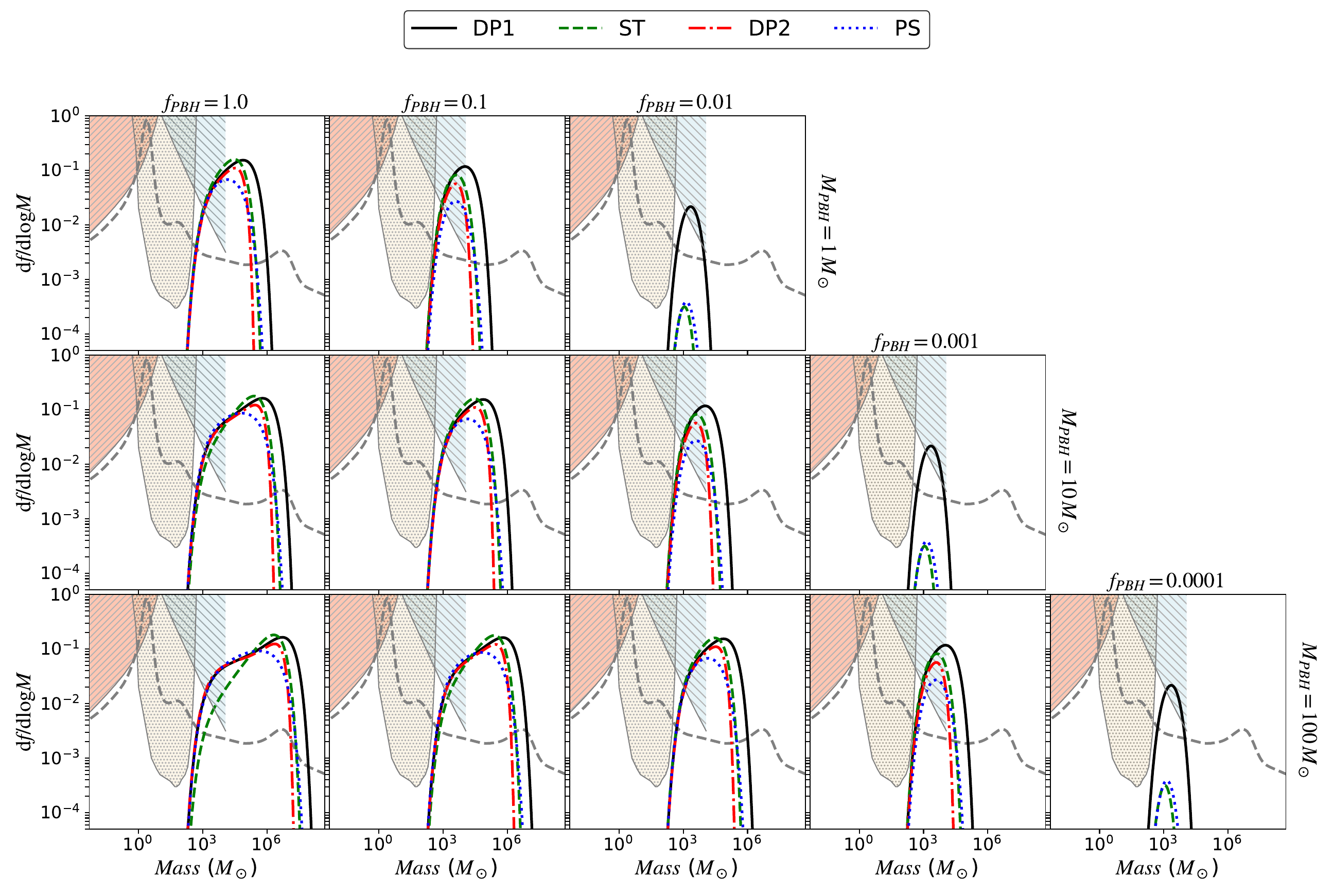}
\caption{Similar to Fig.\,\ref{Fig2} but with $n=2$.}
\label{Fig3}
\end{figure*}

\section{Results and Discussion}\label{sec:iii}
In this section, we analyze and compare the predicted abundance of UDMHs by incorporating modifications to halo formation processes, taking into account more realistic mass functions and the influence of Poisson noise from stellar-mass PBHs. 

In Fig.\,\ref{Fig2}, we have illustrated how the mass of PBHs dramatically shapes the differential mass fraction ${\rm d}f/{\rm d}\log M=(M/\bar{\rho}_{m,0})({\rm d}n/{\rm d}\log M)$ of UDMHs. The panels indicate that as the PBH mass increases from $1 M_\odot$ to $100 M_\odot$, the overall abundance of UDMHs shifts, and their mass distribution becomes increasingly sensitive to the Poissonian isocurvature noise induced by these black holes. This behavior stems from the inverse relation between PBH mass and number density: heavier PBHs generate stronger shot noise due to their lower abundance, which in turn amplifies small-scale fluctuations more effectively, thereby catalyzing the formation of ultradense structures at earlier times. In these panels, the suppression parameter is fixed at $n=1$, corresponding to a relatively mild attenuation of the Poisson-enhanced fluctuations at very high wavenumbers; as a result, the isocurvature contribution remains significant over a broad range of small scales, maximizing its effect on UDMH formation across the entire mass spectrum.
  
The variation of $f_{\mathrm{PBH}}$ from unity down to $10^{-4}$ clearly modulates the strength of the Poisson-induced isocurvature component in the power spectrum, as reflected in the differential mass functions. For higher values of the fraction of PBHs, i.e., $f_{\rm PBH} \rightarrow 1$, the contribution of shot noise is maximized, resulting in a significantly boosted UDMH abundance over a broad mass range. Conversely, at $f_{\mathrm{PBH}} \lesssim 10^{-3}$, the contribution of PBH-induced Poisson noise becomes smaller, and the resulting halo mass function converges toward the small deviations from the standard adiabatic expectation. We have also represented the PBH mass spectrum derived from the thermal history of the early Universe \citep{2019arXiv190608217C}.

Moreover, we have compared the predictions from different halo mass function models: DP1, DP2, ST, and PS. Notably, the DP1 and DP2 models, which include advanced physical corrections such as angular momentum and dynamical friction, consistently predict a higher UDMH abundance relative to the traditional PS formalism. The ST model, incorporating ellipsoidal collapse dynamics, also shows improved consistency with these physical enhancements. The convergence of these models at low mass scales, and their divergence at high mass, underscores the importance of properly accounting for non-linear collapse physics, particularly in regimes dominated by small-scale enhancements such as those induced by PBHs.

\begin{figure*}
\centering
\includegraphics[width=0.8\textwidth]{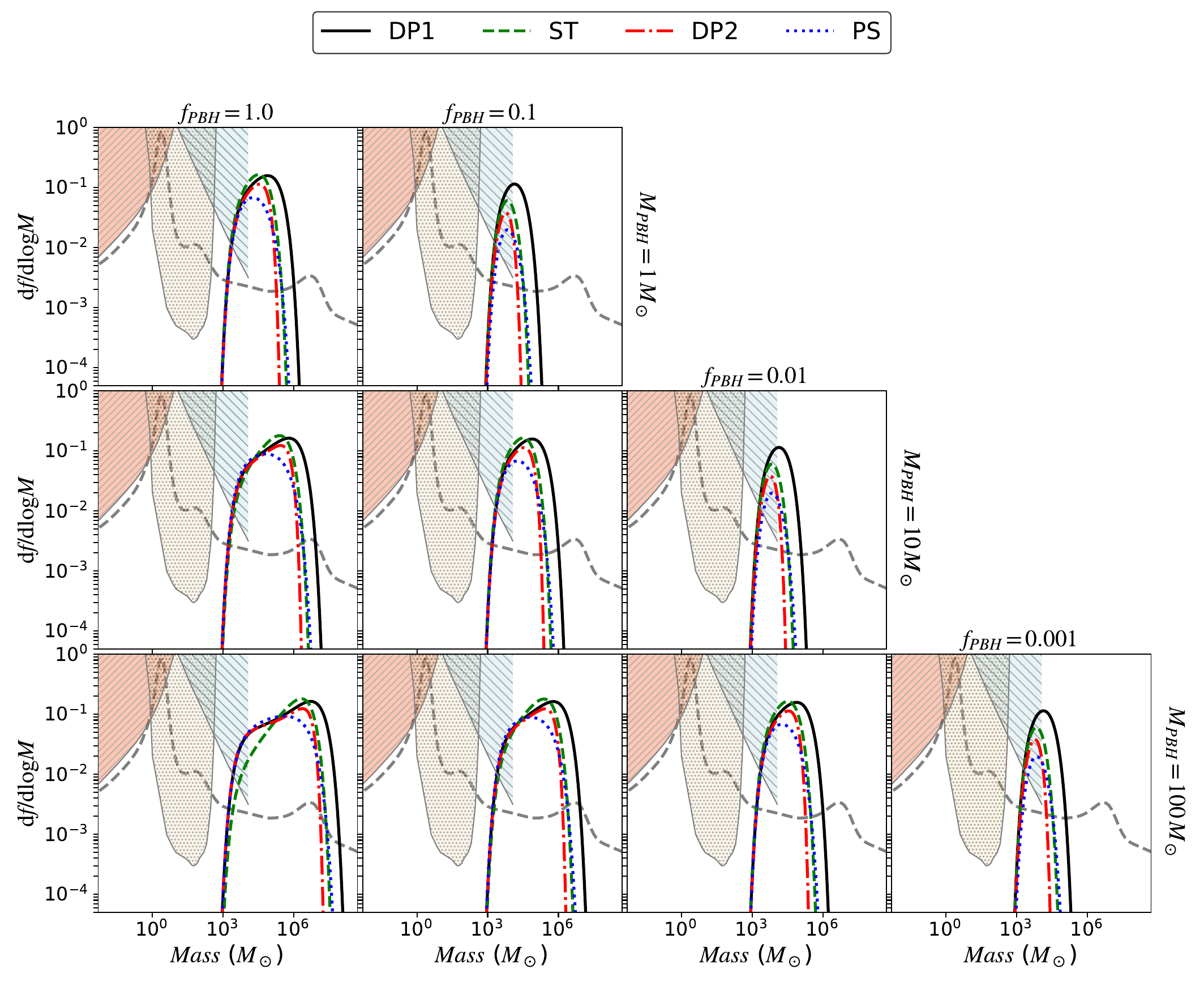}
\caption{Similar to Fig.\,\ref{Fig2} but with $n=3$.}
\label{Fig4}
\end{figure*}

The inclusion of observational limits in Fig.\,\ref{Fig2}, namely from microlensing surveys conducted by OGLE (indicated in shaded blue)  \citep{2024ApJ...976L..19M}, gravitational-wave data from the LVK collaboration (shaded red) \citep{2019PhRvD.100b4017A, 2022PhRvL.129f1104A, 2024PhRvD.110b3040A}, and the dynamical stability of UFD galaxies (shaded green) \citep{2016ApJ...824L..31B}, adds a critical evaluative layer to the model predictions. In almost all panels, the clear violation of regions previously excluded by observational constraints on the abundance of PBHs, suggests that the analysis lends compelling support to a multi-component dark matter scenario. This means that, in addition to the prominent contribution of the Poisson noise effect of primordial black holes, the effects of microscopic dark matter particle clusters should not be neglected. In such cases, the differential mass fraction of UDMHs becomes dominant in a specific mass window that coincides with regions of observational interest, including UFD disruption limits and LVK constraints. The substantial enhancement in UDMH formation reflects that the early universe may have contained both cold dark matter particle and macroscopic PBHs each contributing to structure formation through distinct pathways.

This result highlights that multi-component dark matter scenarios are favored in certain regions of parameter space, particularly when Poisson noise from stellar-mass PBHs enhances the formation of UDMHs. This implies a possible coexistence of PBHs and particle dark matter, where each component contributes to structure formation through distinct mechanisms. Such a scenario opens up new phenomenological possibilities, including unique signatures in gravitational lensing, dynamical friction effects, or indirect detection signals from dark matter annihilation within UDMHs. While the current analysis focuses on observational bounds from microlensing, gravitational wave data, and ultra-faint dwarf galaxy survival, additional constraints, such as those from the Lyman-$\alpha$ forest, could further restrict the viable parameter space for multi-component models \citep[e.g.,][]{2019PhRvL.123g1102M, 2025arXiv250406367G}. These constraints, which probe small-scale matter distribution and suppress power in the linear regime, are especially relevant for assessing the viability of mixed dark matter scenarios.

In Figs.\,\ref{Fig3} and \ref{Fig4}, we have extended the key results from Fig.\,\ref{Fig2} by exploring the impact of increasing the suppression parameter $n$ to 2 and 3, respectively, in the modified power spectrum. This parameter governs the sharpness of the exponential cutoff in the isocurvature (Poisson-induced) fluctuations contributed by stellar-mass PBHs. The figures demonstrate how varying the spectral suppression shape modulates the resultant formation of UDMHs. Both figures systematically explore a parameter space involving same PBH mass and fraction.

Comparing Figs.\,\ref{Fig3}, $(n = 2)$, and \ref{Fig4}, $(n = 3)$, with the earlier Fig.\,\ref{Fig2}, we observe that increasing suppression parameter leads to stronger suppression of Poisson noise at higher wavenumbers. This sharper cutoff significantly reduces the small-scale enhancement in the power spectrum, thereby dampening the formation of UDMHs in the low-mass regime. Particularly, for lower values of $f_{\mathrm{PBH}}$, the abundance curves start to converge toward the predictions from the small deviations from the standard adiabatic spectrum. This demonstrates the delicate balance between small-scale amplification and suppression: while Poisson noise from massive PBHs initially inflates the UDMH mass function, its cosmological influence is ultimately curtailed by how rapidly the cutoff in $P_{\mathrm{iso}}(k)$ declines with $k$.

In both figures, increasing $M_{\mathrm{PBH}}$ shifts the differential mass function ${\rm d}f/{\rm d}\log M$ to higher masses, as expected due to stronger Poisson fluctuations induced by fewer, more massive PBHs. However, the suppression introduced by $n = 2$ and $n = 3$ tempers this effect. At higher $f_{\mathrm{PBH}}$, the noise amplitude $A_{\mathrm{iso}}$ in the power spectrum increases linearly, magnifying the isocurvature impact. Thus, while $f_{\mathrm{PBH}} \rightarrow 1$ results in noticeable boosts in UDMH formation, the increment becomes less prominent as $n$ increases, particularly for lighter PBHs. This reflects a saturation effect, higher-order suppression trims the contributions from smaller scales before they can fully translate into compact structure formation.

As in earlier figures, the DP1 and DP2 mass functions, which incorporate corrections for angular momentum, and dynamical friction, yield the highest UDMH abundances, especially in the high-mass tail. In contrast, the PS formalism underestimates these abundances due to its oversimplified spherical-collapse assumption. The ST mass function falls between DP1 and DP2 predictions and exhibits behavior consistent with Fig.\,\ref{Fig2}.

As illustrated in Fig.\,\ref{Fig3}, adopting a suppression parameter of $n = 2$ yields a differential mass function that exhibits improved agreement with observational data, particularly those from the LVK collaboration. This adjustment enhances the plausibility of a single-component dark matter model in which UDMHs can be formed by lighter PBHs. In comparison to the $n = 1$ case, the results suggest a more consistent alignment with the observation constraints on the abundance of PBHs. Nevertheless, a mild tension remains with the upper bounds inferred from the survival of UFDs. These outcomes are compatible with the PBH mass spectrum shaped by the thermal history of the early Universe and underscore the differentiated role played by PBHs of varying masses in the formation of UDMHs.

Moving to Fig.\,\ref{Fig4}, one can observe that higher suppression values further reinforce the viability of a single-component dark matter scenario. The predicted abundance of UDMHs deviate less from observational constraints on the abundance of PBHs. Notably, the resulting mass spectra display closer consistency with those anticipated from thermal models of early-universe evolution, wherein lighter PBHs emerge as the dominant contributors to the dark matter content. This alignment suggests a coherent narrative in which small-scale enhancements, shaped by Poisson fluctuations, favor early formation of UDMHs, driven by quantum-origin PBHs in the stellar mass regime.
\section{Conclusions}\label{sec:iv}
In this study, we have thoroughly investigated the impact of Poisson noise generated by stellar-mass PBHs on the formation and abundance of UDMHs. By incorporating shot noise contributions into the power spectrum, we have demonstrated that the discrete distribution of PBHs significantly alters small-scale density fluctuations, subsequently enhancing the formation of compact dark matter structures during the radiation-dominated era. Our modified matter power spectrum, which accounts for both adiabatic perturbations and isocurvature contributions from PBH Poisson statistics, reveals that the effectiveness of this enhancement is strongly dependent on the mass of the contributing PBHs and their relative abundance in the dark matter content.

The mass of PBHs emerges as a critical parameter governing the properties of resulting UDMHs. Our analysis shows that heavier PBHs $(10\mbox{-}100 M_\odot)$ generate stronger shot noise due to their naturally lower number density, leading to more pronounced enhancements in small-scale power spectrum. This effect produces a distinctive signature in the differential mass fraction of UDMHs, shifting their abundance toward higher masses as $M_{\rm PBH}$ increases. Conversely, lighter PBHs (around $1 M_{\odot}$) induce more moderate modifications to the power spectrum, resulting in UDMH distributions that remain closer to small deviations from standard adiabatic fluctuations, particularly at lower $f_{\rm PBH}$ values and higher suppression parameters.

The suppression parameter $n$ in our modified power spectrum serves as an essential regulatory mechanism for the influence of Poisson noise on UDMH formation. Our results demonstrate that increasing $n$ from $1$ to $3$ progressively dampens the contribution of shot noise at higher wavenumbers, effectively reducing the formation efficiency of lower-mass UDMHs. This suppression creates varying degrees of alignment with observational constraints, with higher $n$ values generally producing UDMH abundances more consistent with existing bounds from microlensing surveys conducted by OGLE, gravitational-wave data from the LVK collaboration, and the dynamical stability of UFD galaxies. Importantly, we find that increasing $n$ leads to results more consistent with current observational constraints, indicating that sharper cutoffs may better reflect realistic suppression of small-scale power in PBH-dominated scenarios.

The comparison of multiple halo mass function formalisms reveals the importance of incorporating more realistic physical processes in modeling UDMH formation. The DP1 and DP2 models, which account for angular momentum and dynamical friction effects, consistently predict higher UDMH abundances compared to the traditional PS approach, particularly in the high-mass regime. These enhancements underscore the necessity of moving beyond simplistic spherical collapse approximations when examining structure formation triggered by non-standard processes such as PBH-induced Poisson fluctuations.

When evaluated against observational constraints, our findings suggest a nuanced picture regarding the composition of dark matter. For lower suppression parameters ($n=1$) and higher PBH masses, the predicted UDMH abundances frequently exceed observational bounds, lending support to multi-component dark matter scenarios where both PBHs and particle dark matter contribute significantly. However, as the suppression parameter increases ($n=2,$ and $3$), particularly for lighter PBHs, the results become increasingly compatible with a single-component dark matter model dominated by stellar-mass PBHs, aligning more closely with mass spectra derived from the thermal history of the early Universe.

While our study primarily employs an analytic framework to explore the impact of Poisson noise from stellar-mass PBHs on UDMH formation, we acknowledge the value of numerical validation, especially for small-scale fluctuations. The modified power spectrum and mass functions used in our analysis are based on well-established theoretical models that have been tested against simulations in prior studies. These include the excursion set formalism with moving barriers and ellipsoidal collapse dynamics, which have shown robustness across different cosmological scenarios. Our approach ensures convergence through careful sampling of mass ranges and wavenumber integrations. Nonetheless, future high-resolution $N$-body simulations will be important to further validate our results and capture potential nonlinear effects not included in our current treatment.

Finally, the viability of PBH-dominated dark matter scenarios is evaluated using current observational constraints, such as microlensing and gravitational wave data, which carry uncertainties related to PBH mass distribution, clustering, and detection efficiency. These uncertainties affect the robustness of the derived bounds and, consequently, the compatibility of PBH models with observations. Our findings suggest that PBH-induced Poisson fluctuations can produce UDMHs with structural features distinct from $\Lambda$CDM predictions. These signatures may be testable by future gravitational lensing surveys or indirect detection methods. However, refining PBH formation models, small-scale power spectra, and nonlinear clustering through simulations and upcoming observations will be essential for validating these predictions and constraining the role of Poisson noise in dark matter structure formation.


\bibliography{sample7}{}
\bibliographystyle{aasjournalv7}



\end{document}